\documentclass{article}

\usepackage{microtype}
\usepackage{graphicx}
\usepackage{subcaption}
\usepackage{booktabs} 

\usepackage{hyperref}

\usepackage[accepted]{icml2025}

\usepackage{amsmath}
\usepackage{amssymb}
\usepackage{mathtools}
\usepackage{amsthm}
\usepackage{cancel}
\usepackage{bm}

\usepackage[capitalize,noabbrev]{cleveref}

\definecolor{myred}{RGB}{228,26,28}
\definecolor{myblue}{RGB}{55,126,183}
\definecolor{mygreen}{RGB}{77,174,74}

\theoremstyle{plain}

\theoremstyle{definition}

\theoremstyle{remark}

\usepackage[textsize=tiny]{todonotes}

\icmltitlerunning{Multi-Fidelity SBI}

\begin{document}

\twocolumn[
\icmltitle{Simulation-Efficient Cosmological Inference with Multi-Fidelity SBI}



\icmlsetsymbol{equal}{*}

\begin{icmlauthorlist}
\icmlauthor{Leander Thiele}{ipmu,cd3}
\icmlauthor{Adrian E. Bayer}{cca,princ}
\icmlauthor{Naoya Takeishi}{rcast}
\end{icmlauthorlist}

\icmlaffiliation{ipmu}{Kavli IPMU (WPI), UTIAS, The University of Tokyo, Japan}
\icmlaffiliation{cd3}{Center for Data-Driven Discovery, Kavli IPMU (WPI), Japan}
\icmlaffiliation{cca}{Center for Computational Astrophysics, Flatiron Institute, New York, USA}
\icmlaffiliation{princ}{Department of Astrophysical Sciences, Princeton University, USA}
\icmlaffiliation{rcast}{RCAST, The University of Tokyo, Japan}

\icmlcorrespondingauthor{Leander Thiele}{leander.thiele@ipmu.jp}

\icmlkeywords{Simulation-Based Inference, Multi-Fidelity Simulation}

\vskip 0.3in
]



\printAffiliationsAndNotice{}  

\begin{abstract}
The simulation cost for cosmological simulation-based inference can be decreased by combining simulation sets of varying fidelity.
We propose an approach to such multi-fidelity inference based on feature matching and knowledge distillation.
Our method results in improved posterior quality, particularly for small simulation budgets and difficult inference problems.
\end{abstract}

\section{Introduction}
\label{introduction}

Cosmology is seeing an increase in attention for simulation-based inference (SBI) methods.
Many proofs-of-concept have appeared, and even applications to real data are now possible~\citep[e.g.,][]{Hahn2024,Lemos2024,Gatti2024,Thiele2024,Novaes2025}.
Such applications do not, at the moment, enjoy the same degree of trust as do traditional likelihood-based analyses.
To increase the trustworthiness of SBI in cosmology methodological developments are necessary,
in particular in maximizing the posterior's quality with a realistic simulation budget.

SBI enables us to constrain model parameters in situations where the likelihood of an
observed data vector is not tractable.
In many such cases, we are able to simulate samples from the likelihood.
Given a training set of such simulated samples, SBI~\cite{Cranmer2020} is an umbrella
for algorithms that can learn approximations to the likelihood or functionally equivalent objects: e.g.,
posterior estimation~\citep[NPE,][]{Greenberg2019,Lueckmann2027,Papamakarios2016},
likelihood estimation~\citep[NLE,][]{Papamakarios2019nle,Lueckmann2019nle},
ratio estimation~\citep[NRE,][]{Hermans2020nre},
and quantile estimation~\citep[NQE,][]{Jia2024a}.

There is no free lunch in SBI. Posterior approximations can be dramatically wrong~\citep{Hermans2022}.
This is particularly pertinent in situations with a very limited simulation budget~\citep{Lueckmann2021,Homer2024,Bairagi2025}.
Cosmological simulations are extremely expensive if they are to capture all relevant physics accurately.
For current and upcoming data sets, we will only be able to run very few simulations with a fidelity matching the data precision.
This problem has been studied in a few recent works.
By combining accurate analytic descriptions on large scales with cheaper small-scale simulations,
the computational expense can be decreased~\citep{modi2023hybridsbiilearned,Zhang2025}.
Sequential methods~\citep[e.g.,][]{Cole2022} concentrate the simulation budget in the most informative regions of parameter space.

\begin{figure}[!t]
    \centering
    \includegraphics[width=\linewidth]{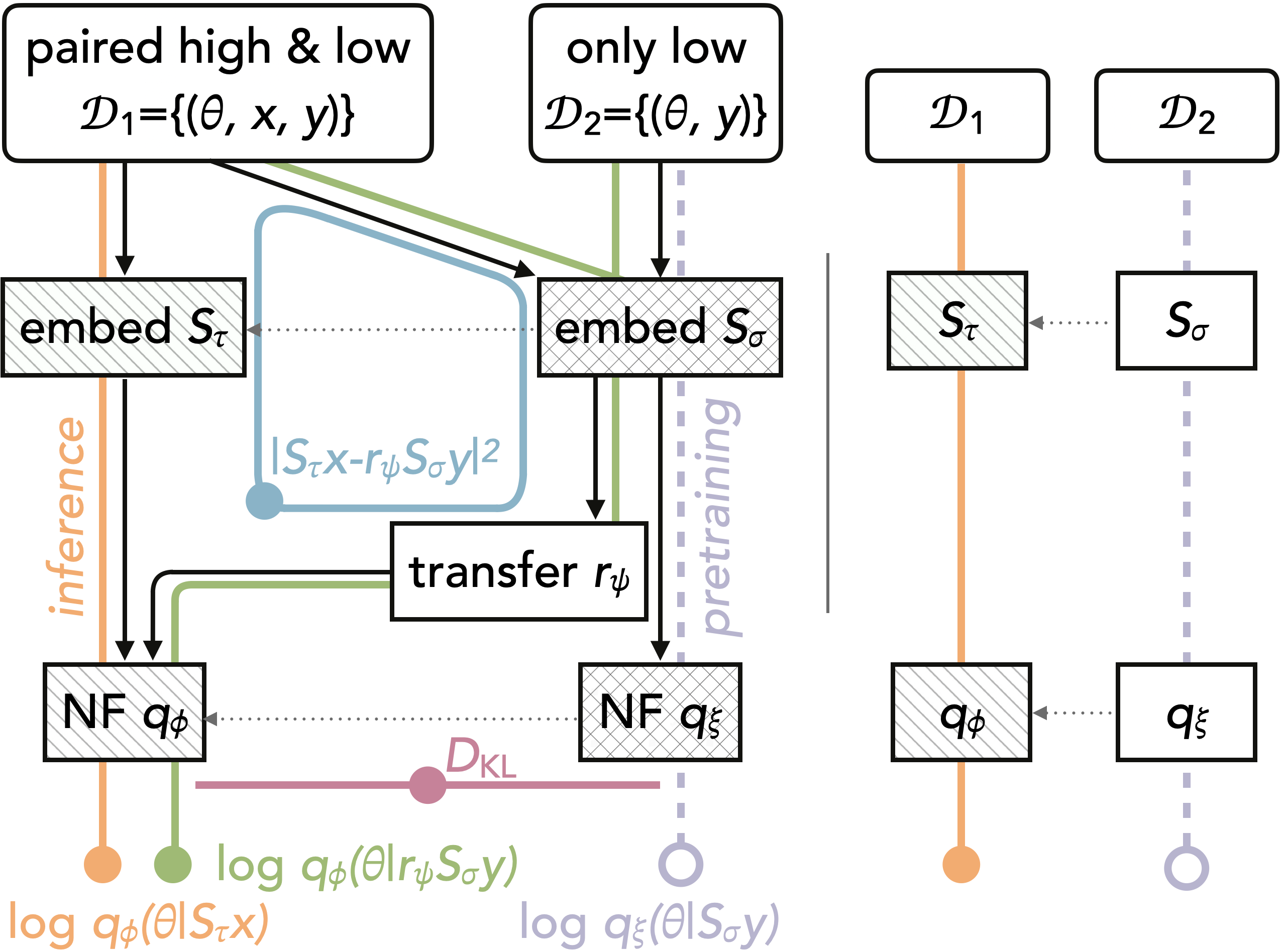}
    \caption{Graphical summary of the proposed multi-fidelity architecture and training loss. Filled circles indicate the components of our proposed training loss.
    The notation is given in \cref{methods}.
    Hatched components are weight-initialized (dotted arrows) from pretraining on $\mathcal{D}_2$, the cross-hatched ones being frozen.
    The final model only evaluates the path indicated in orange.
    The right side shows the weight-initialization scheme we compare to.
    }
    \label{fig:schema}
\end{figure}

\begin{figure*}[t!]
    \centering
    \includegraphics[width=0.33\linewidth]{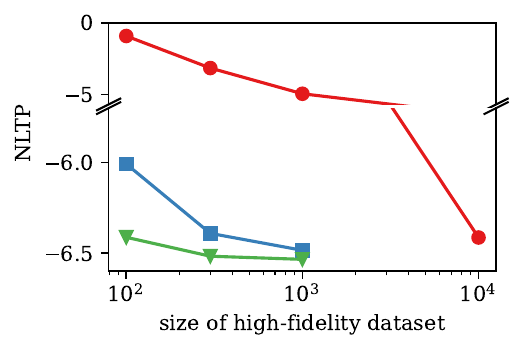}
    \includegraphics[width=0.33\linewidth]{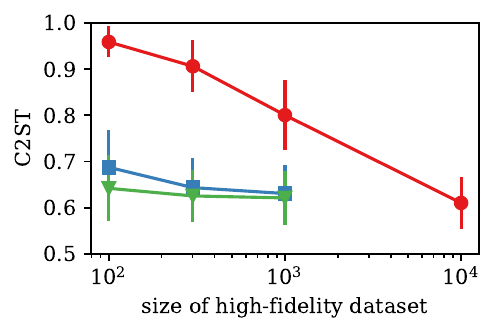}
    \includegraphics[width=0.33\linewidth]{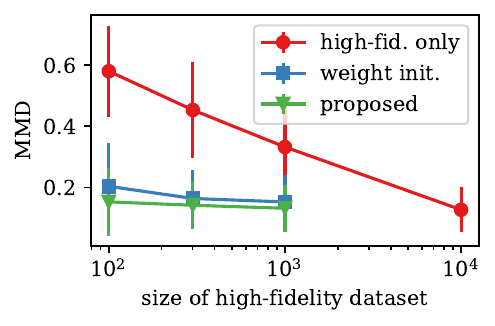}
    \vspace*{-2em}
    \caption{Performance evaluation according to various metrics for the cosmological inference problem. Lower is better for all.}
    \label{fig:eval_cosmo}
\end{figure*}

In this work, we focus on the approach of multi-fidelity inference.
In this setup, we assume a cascade of simulation sets with varying levels of fidelity,
culminating in a small high-fidelity set matching the observation's precision.
Prior work in this direction, specializing on the case of two fidelity levels, has proposed a calibration step in NQE~\citep{Jia2024},
and transfer learning via weight initialization~\citep{Krouglova2025,Saoulis2025}.
These existing approaches highlight the potential of multi-fidelity SBI to reduce the required simulation budget.

Our contribution employs a tailored training setup which enables training on multiple fidelity simulations simultaneously.
The training constructs stochastic mappings between the embedded data vectors at different fidelity levels
and a latent space corresponding to the highest fidelity.
\cref{fig:schema} illustrates our architecture and training process, a derivation is presented in \cref{methods}.
Our approach is a superset of weight initialization.
We demonstrate that it outperforms weight initialization in examples where the two can be compared.
Furthermore, our approach can accommodate any number of fidelity levels
and is applicable even in situations when the observations or embeddings at different fidelities differ in shape.
Empirically, it is also found to converge faster than weight initialization.

In this work, we focus on the case of NPE.
The extension to NRE appears natural but will be deferred to future work.
We consider a cosmology example with a traditional summary statistic, though the extension to field-level inference, as considered in \citet{Saoulis2025}, is straightforward.


\section{Results}
\label{results}

We evaluate performance by considering the example of the matter power spectrum at redshift $z=0$ in 5-parameter $\Lambda$CDM.
The high-fidelity data is taken from the BSQ set of the Quijote simulation suite~\citep{Villaescusa-Navarro_2020}, which uses a tree-PM algorithm for accurate small scale clustering.
The low-fidelity data was produced using FastPM~\citep{Feng2016}, a PM code which is cheaper by a factor of $\sim$10--100 per simulation, using the same prior.
The FastPM simulations paired with Quijote ($\mathcal{D}_1$) are matched in parameters and seeds.\footnote{Matching seeds is, in principle, not required as the transfer network $r_\psi$ can absorb the additional stochasticity. However, performance of the proposed method benefits from matching the seeds.}
We truncate at $k_\text{max}=0.5\,h\text{Mpc}^{-1}$ to dimensionality 79.

We take MLPs for the embedding and transfer networks, and a spline flow~\citep{Durkan2019} from \texttt{sbi}~\citep{tejero-cantero2020sbi,nflows} for the density estimator.
Hyperparameters are optimized with \texttt{optuna}~\citep{akiba2019}.

For all multi-fidelity runs we use 10k low-fidelity samples with  100, 300, or 1k high-fidelity samples.
We evaluate performance using a held-out test set of 2k high-fidelity simulations,
and compare to reference posteriors computed by training standard NPE on 28k high-fidelity samples.

\begin{figure}[!t]
    \centering
    \includegraphics[width=\linewidth]{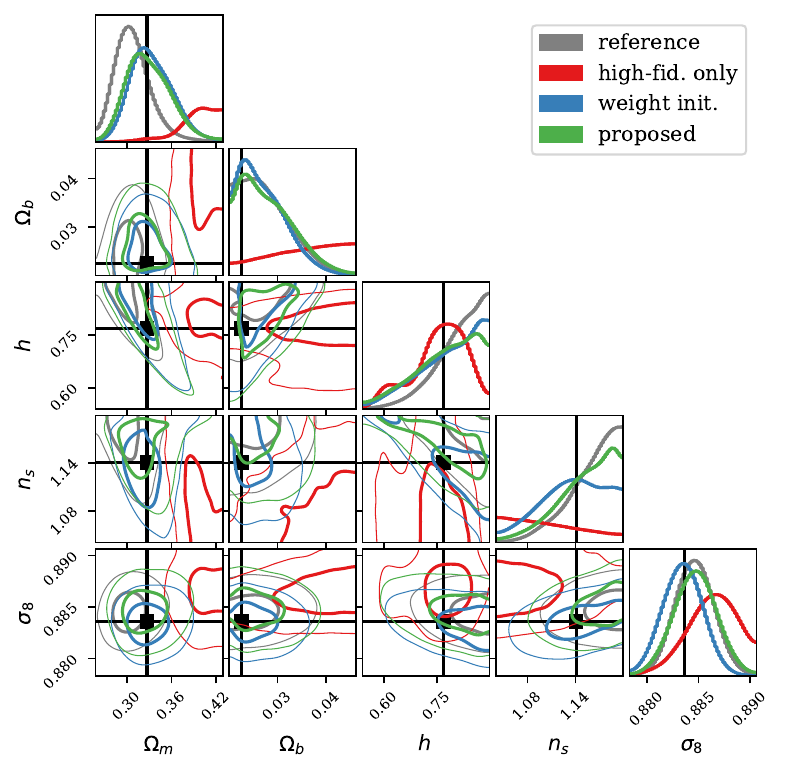}
    \vspace*{-2em}
    \caption{Posteriors on an example realization from the test set. We compare models trained on 100 high-fidelity simulations to the reference posterior (grey) trained on 28k high-fidelity simulations.}
    \label{fig:corner}
\end{figure}

\cref{fig:eval_cosmo} shows three different evaluation metrics as a function of the high-fidelity set size.
NLTP is the test loss, C2ST is the two-sample classifier accuracy~\citep{lopezpaz2018}, and MMD is the kernel-based maximum mean discrepancy~\citep{gretton12a}.
\textcolor{mygreen}{Our approach} consistently outperforms \textcolor{myblue}{weight initialization}, with most benefit for small high-fidelity sets.
As an example, in \cref{fig:corner} we show a comparison of posteriors.
\textcolor{mygreen}{Our method} results in posteriors quite close to those obtained with \textcolor{myblue}{weight initialization},
while the estimation \textcolor{myred}{only with small high-fidelity data} is far from the reference.



\section{Methods}
\label{methods}

\subsection{Preliminaries}

NPE is a problem to estimate the conditional distribution $p(\theta \mid x)$ given realizations of simulator's parameters $\theta$ and outcomes $x$ as training data.
In the standard NPE, we learn a density estimator $q_\phi(\theta \mid x)$ parametrized by $\phi$, typically a conditional normalizing flow, by minimizing
\begin{multline*}
    \mathbb{E}_{p(x)} \operatorname{KL} \big( p(\theta \mid x) \,||\, q_\phi(\theta \mid x) \big)
    \\
    = - \mathbb{E}_{p(\theta, x)} \log q_\phi(\theta \mid x) - \mathbb{E}_{p(x)} H(p(\theta \mid x)).
\end{multline*}
Let $\mathcal{D}=\{(\theta_1, x_1), \dots, (\theta_n, x_n)\}$ be data drawn from $p(x \mid \theta)p(\theta)$.
As the entropy $H(p)$ is constant wrt $\phi$, we solve
\begin{equation}\label{eq:npeloss}
    \operatorname*{minimize}_\phi ~~ \frac1n \sum_{(\theta,x) \in \mathcal{D}} -\log q_\phi(\theta \mid x).
\end{equation}

\subsection{Proposed Method}

Let $p(x \mid \theta)$ and $p(y \mid \theta)$ be the likelihoods of two simulators to model the same target with different fidelity: $x \in \mathcal{X}$ is with relatively high fidelity, and $y \in \mathcal{Y}$ is with lower fidelity.
Evaluating the values of $p(x \mid \theta)$ and $p(y \mid \theta)$ is intractable, and we only have samples drawn from them.
The high-fidelity simulator is much heavier to run than the low-fidelity simulator, so we typically have $\mathcal{D}_1=\{(\theta_i, x_i, y_i) \mid i=1,\dots,n_1\}$ and $\mathcal{D}_2=\{(\theta_i, y_i) \mid i=1,\dots,n_2\}$, where $n_1 < n_2$ by a large margin.

Our goal is to estimate the conditional distribution of $\theta \mid x$.
However, the standard NPE loss in \cref{eq:npeloss} for $x$ can be computed only on the $\theta$-$x$ pairs in small $\mathcal{D}_1$.
We thus want to utilize the information from the $\theta$-$y$ pairs, available in larger $\mathcal{D}_2$ as well, for learning the $\theta$-$x$ relation.
To this end, we introduce additional loss functions as presented in \cref{newloss1,newloss2}.
For the sake of simplicity, we here only discuss the case of two levels of fidelity, but extending the method to more fidelity levels is straightforward.

\subsubsection{Feature matching}
\label{newloss1}

We aim to minimize the $y$-based posterior KL:
\begin{multline*}
\mathbb{E}_{p(y)} \operatorname{KL} \big( p(\theta \mid y) \,||\, q_\phi(\theta \mid y) \big)
\\
= -\mathbb{E}_{p(\theta,y)} \log q_\phi(\theta \mid y) - \mathbb{E}_{p(y)} H(p(\theta \mid y)).
\end{multline*}
Recall that the network of the density estimator, $q_\phi$, is designed to receive $x$ (and not $y$) as the condition, so we cannot compute $-\log q_\phi(\theta \mid y)$ directly.
It can be upper bounded by Jensen's inequality as
\begin{equation*}\begin{aligned}
& -\log q_\phi(\theta \mid y)
= -\log \int r_\psi(x \mid y) \frac{q_\phi(\theta \mid x) p(x \mid y)}{r_\psi(x \mid y)} dx
\\ &\quad
\leq -\int r_\psi(x \mid y) \log \frac{q_\phi(\theta \mid x) p(x \mid y)}{r_\psi(x \mid y)} dx
\\ &\quad
= - \mathbb{E}_{r_\psi(x \mid y)} \log q_\phi(\theta \mid x) + \operatorname{KL} \big( r_\psi(x \mid y) \ || \ p(x \mid y) \big),
\end{aligned}\end{equation*}
where $r_\psi(x \mid y)$ is a conditional distribution of $x$ given $y$ parametrized with $\psi$.
We thus should minimize
\begin{gather}
    - \mathbb{E}_{p(\theta, y)} \mathbb{E}_{r_\psi(x \mid y)} \log q_\phi(\theta \mid x) \quad\text{and} \label{eq:newloss1}
    \\
    \mathbb{E}_{p(y)} \operatorname{KL} \big( r_\psi(x \mid y) \ || \ p(x \mid y) \big). \label{eq:newloss2}
\end{gather}

The newly introduced model, $r_\psi(x \mid y)$, probabilistically transforms $y$ to $x$.\footnote{We will discuss in \cref{implementation} more practical cases where embedding nets are applied to $x$ and $y$.}
\Cref{eq:newloss1} reads as the NPE loss computed on the transformed condition $x \sim r_\psi(x \mid y)$.
\Cref{eq:newloss2} enforces $r_\psi(x \mid y)$ close to the true transformation, but computing the KL is challenging because we cannot evaluate $p(x \mid y)$ and only have access to the samples from $p(x, y)$.
As a surrogate of \eqref{eq:newloss2}, we suggest minimizing the KL in the opposite direction:
\begin{multline}\label{eq:newloss2_reverse}
    \tag{\ref*{eq:newloss2}$'$}
    \mathbb{E}_{p(y)} \operatorname{KL} \big( p(x \mid y) \ || \ r_\psi(x \mid y) \big)
    \\ = - \mathbb{E}_{p(x,y)} \log r_\psi(x \mid y) + \text{const}.
\end{multline}

\subsubsection{Response distillation}
\label{newloss2}

Let $\tilde{q}_\xi(\theta \mid y)$ be the density estimator of $p(\theta \mid y)$ trained on the large dataset $\mathcal{D}_2$ with the standard NPE loss, i.e.,
\begin{equation*}
    \xi = \arg\min_{\xi'} ~~ \frac1{n_2} \sum_{(\theta,y) \in \mathcal{D}_2} -\log \tilde{q}_{\xi'}(\theta \mid y).
\end{equation*}
We aim to inform our estimator, $q_\phi(\theta \mid x)$, from $\tilde{q}_\xi(\theta \mid y)$ based on knowledge distillation \citep{hintonDistillingKnowledgeNeural2015}.
We minimize the (expectation of) KL between the two:
\begin{multline*}
\mathbb{E}_{p(y)} \operatorname{KL}(\tilde{q}_\xi(\theta \mid y) \ || \ q_\phi(\theta \mid y))
\\
= - \mathbb{E}_{p(y)} \mathbb{E}_{\tilde{q}_\xi(\theta \mid y)} \log q_\phi(\theta \mid y) - \mathbb{E}_{p(y)} H(\tilde{q}_\xi(\theta \mid y)).
\end{multline*}
Since $\xi$ is fixed already, the last term is constant.
As $-\log q_\phi(\theta \mid y)$ can be upper bounded as discussed earlier, the quantity to minimize is
\begin{multline*}
- \mathbb{E}_{p(y)} \mathbb{E}_{\tilde{q}_\xi(\theta \mid y)} \log q_\phi(\theta \mid y)
\\
\leq - \mathbb{E}_{p(y)} \mathbb{E}_{\tilde{q}_\xi(\theta \mid y)} \mathbb{E}_{r_\psi(x \mid y)} \log q_\phi(\theta \mid x)
\\
+ \underbrace{\mathbb{E}_{p(y)} \operatorname{KL}(r_\psi(x \mid y) \ || \ p(x \mid y))}_\text{same as \eqref{eq:newloss2}}.
\end{multline*}
Note that the KL in the last term already appeared in \cref{eq:newloss2}.
Finally we are interested in minimizing
\begin{equation}
- \mathbb{E}_{p(y)} \mathbb{E}_{\tilde{q}_\xi(\theta \mid y)} \mathbb{E}_{r_\psi(x \mid y)} \log q_\phi(\theta \mid x),
\end{equation}
which is equivalent to minimizing the KL divergence between $\tilde{q}_\xi(\theta \mid y)$ and $q_\phi(\theta \mid x)$.

\subsubsection{Implementation}
\label{implementation}

We often use so-called embedding networks to extract summary statistics from observations, which were omitted in the discussion so far for notational simplicity.
Let $S_\tau: x \mapsto S_\tau(x)$ and $S_\sigma: y \mapsto S_\sigma(y)$ be the embedding networks for $x$ and $y$, parametrized with $\tau$ and $\sigma$, respectively.
With these embedding nets taken into consideration, the proposed method proceeds as follows.

\paragraph{Step 1.}
Train an NPE model on $\mathcal{D}_2$ ordinarily by
\begin{equation*}
    \operatorname*{minimize}_{\xi,\sigma} ~~ \frac1{n_2} \sum_{(\theta,y) \in \mathcal{D}_2} -\log \tilde{q}_\xi(\theta \mid S_\sigma(y)).
\end{equation*}

\paragraph{Step 2.}
Train an NPE model $q_\phi(\theta \mid S_\tau(x))$ with the proposed loss.
First, if possible, initialize $\phi$ with $\xi$.
We then solve the following problem:
\begin{equation}
    \operatorname*{minimize}_{\phi, \tau, \psi} ~~ L_1 + \alpha L_2 + \beta L_3,
\end{equation}
where each loss term is given as
\begin{equation*}\scalebox{0.9}{$\begin{aligned}
    L_1 &= \frac1{n_1} \sum_{(\theta, x) \in \mathcal{D}_1} -\log q_\phi (\theta \mid S_\tau(x)),
    \\
    L_2 &= \frac1{n_{1,2}} \sum_{(\theta, y) \in \mathcal{D}_{1,2}} \mathbb{E}_{u \sim r_\psi(\cdot \mid S_\sigma(y))} -\log q_\phi(\theta \mid u) \\
    &\qquad\qquad\qquad + \frac1{n_1} \sum_{(x, y) \in \mathcal{D}_1} -\log r_\psi (S_\tau(x) \mid S_\sigma(y)),
    \\
    L_3 &= \frac1{n_{1,2}} \sum_{y \in \mathcal{D}_{1,2}} \mathbb{E}_{\theta \sim \tilde{q}_\xi(\theta \mid S_\sigma(y)), \ u \sim r_\psi(\cdot \mid S_\sigma(y))} -\log q_\phi (\theta \mid u).
\end{aligned}$}\end{equation*}
Here, $\mathcal{D}_{1,2} = \mathcal{D}_1 \cup \mathcal{D}_2$ and $n_{1,2} = n_1 + n_2$.
The expectations can be approximated with samples.
We empirically found that instead of rigorously computing $-\log r_\psi (S_\tau(x) \mid S_\sigma(y))$, merely minimizing the squared error $\mathbb{E}_v \Vert S_\tau(x) - v \Vert^2$ where $v \sim r_\psi(\cdot \mid S_\sigma (y))$ resulted in more stable learning.
We note that further analysis of this heuristic may lead to an interesting discussion.

\section{Some More Numerical Examples}

\begin{figure}[t]
    \centering
    \begin{minipage}[t]{0.495\linewidth}
        \centering
        \vspace*{0pt}
        \includegraphics[width=\linewidth]{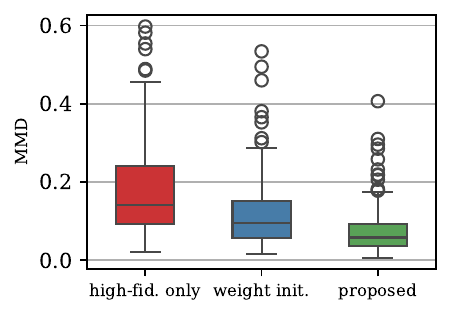}
    \end{minipage}
    \hfill
    \begin{minipage}[t]{0.495\linewidth}
        \centering
        \vspace*{0pt}
        \includegraphics[width=\linewidth]{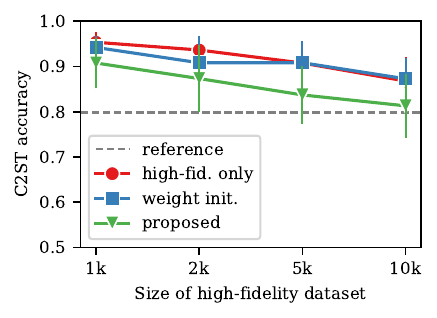}
    \end{minipage}
    \caption{(\emph{Left}) MMD between the true and learned posteriors for \textsc{Gaussian}. (\emph{Right}) C2ST metrics for \textsc{SLCP}. The reference performance is by NPE on a large high-fid. set with $n_1=100\mathrm{k}$.}
    \label{fig:examples}
\end{figure}

For controlled evaluation, we applied the proposed method to two synthetic tasks: \textsc{Gaussian} and \textsc{SLCP}.

In \textsc{Gaussian}, $\theta$ is drawn from the standard normal distribution, and $x$ and $y$ are given as the affine transformation of $\theta$ plus Gaussian noise.
$y$ is supposed to have ``low-fidelity'' by setting $\dim y < \dim x$ and being added stronger noise than $x$ is.
We used $\dim \theta=10$, $\dim x = 10$, and $\dim y = 5$.
We prepared a high-fidelity dataset of size $n_1=100$ and a low-fidelity dataset of size $n_2=5\mathrm{k}$.

\textsc{SLCP} is a standard task in SBI benchmarking \citep[see, e.g.,][]{Lueckmann2021}, featuring multimodal posteriors.
Originally the task is to infer $5$ parameters of a 2D Gaussian distribution given $4$ samples drawn from it.
We made a ``low-fidelity'' version by taking only $3$ samples instead.
So for \textsc{SLCP}, $\dim \theta = 5$, $\dim x = 8$, and $\dim y = 6$.
We tried different sizes of high-fidelity dataset, $n_1=1\mathrm{k}$, $2\mathrm{k}$, $5\mathrm{k}$, or $10\mathrm{k}$, and used the fixed size of low-fidelity dataset with $n_2=100\mathrm{k}$.
To provide a reference performance, we also run an ordinary NPE on a large high-fidelity dataset of size $n_1=100\mathrm{k}$.

The results are shown in \cref{fig:examples}.
We compare the \textcolor{mygreen}{proposed method} with two baselines: models trained only on the $\theta$-$x$ pairs in $\mathcal{D}_1$ (\textcolor{myred}{high-fid. only}) and models pretrained on $\mathcal{D}_2$ and then finetuned on $\mathcal{D}_1$ (\textcolor{myblue}{weight init.}).
For the \textsc{Gaussian} task, the MMD between the true and the learned posterior distributions are reported in the left panel.
For \textsc{SLCP}, the metrics of C2ST are reported in the right panel.\footnote{For \textsc{Gaussian}, MMD should be reliable because the posteriors are unimodal. For \textsc{SLCP}, MMD may be unreliable as the posteriors are multimodal, thus we report C2ST metrics.}
Both sets of results show the efficacy of the proposed method.
It achieves better performance in both tasks than the weight initialization.


\section{Conclusions}

We have presented a new method for multi-fidelity SBI.
Our method contains transfer learning via weight initialization as a special case but extends it in several ways.
Additional loss terms contribute information from lower-fidelity samples during the fine-tuning phase.
Our setup allows for training with an arbitrary number of fidelity levels,
as well as shape mismatch in the data summaries or embeddings.

In our evaluation on a cosmology example problem we find moderate improvement from our method over weight initialization.
The improvement is more pronounced in examples with complicated posterior shape (such as \textsc{SLCP}).
Our proposed method contains several heuristics which could enable improvement in future work.



\section*{Acknowledgements}
We thank S.~Wagner-Carena, O.~Philcox, and C.~Modi for discussions and comments on this draft.
We thank the anonymous reviewers for the ICML-colocated ML4Astro 2025 workshop for their constructive feedback.
This research used computing resources at Kavli IPMU and the Flatiron Institute.
The Kavli IPMU is supported by the WPI (World Premier International Research Center) Initiative of the MEXT (Japanese Ministry of Education, Culture, Sports, Science and Technology).
Leander Thiele was supported by JSPS KAKENHI Grant 24K22878.
Adrian Bayer was supported by the Simons Foundation.
Naoya Takeishi was supported by JSPS KAKENHI Grant Numbers JP20K19869 and JP25H01454 and JSPS International Joint Research Program JPJSJRP20221501.

\bibliography{main}
\bibliographystyle{icml2025}




\end{document}